\documentstyle[12pt,a4wide,epsf]{article}

\begin{document}

\begin{center}
{\large\bf
Search for {\large $\eta$}-mesic nuclei \\ in photoproduction processes}

\bigskip
G.A.~Sokol, T.A.~Aibergenov, A.V.~Kravtsov,\\ A.I.~L'vov, L.N.~Pavlyuchenko

\medskip
{\small \it  P.N. Lebedev Physical Institute of RAS,
  Leninsky Prospect 53, Moscow 117924, Russia}

\end{center}

\begin{abstract}
We present preliminary results of an experiment performed at the
1-GeV electron synchrotron of the Lebedev Physical Institute.  Using the
bremsstrahlung photon beam with the end-point energy of 650--850 MeV
and the carbon target, correlated $\pi^+n$ pairs with opening angle
$\langle\theta_{\pi N}\rangle = 180^\circ$ and energies $\langle
E_{\pi^+}\rangle=300$ MeV, $\langle E_n\rangle=100$ MeV have been
observed. They arise from the process $\gamma + {}^{12}\mbox{C} \to N +
{}_\eta(A-1) \to N + \pi^+ n + (A-2)$ and provide evidence for the
existence of ${}^{11}_{~\eta}$B and ${}^{11}_{~\eta}$C $\eta$-mesic
nuclei.
\end{abstract}

\bigskip
\noindent
{\small PACS numbers:
25.20.Lj, 
24.30.Gd, 
21.30.Fe  
\hfill UDC 539.126}

\vspace{10mm}

Eta-mesic nuclei ${}_\eta A$ are a new sort of nuclear matter which
is bound state of the $\eta$-meson and a nucleus \cite{liu86}.
Early attempts to discover the $\eta$-nuclei in pion-induced
reactions \cite{chr88,lei88} yielded negative results
which excluded originally assumed properties of ${}_\eta A$
\cite{liu86} that were challenged later \cite{chi88,kul98}.
A new interest to study the hypothetical $\eta$-nuclei arose from
an indirect evidence for a formation of the quasi-bound $_\eta A$
state in the reactions $pd \to \eta\, {^3\rm He}$ and
$dd \to \eta \,{^4\rm He}$ which would
naturally explain \cite{wil93,kon94} an experimentally observed
near-threshold enhancement in the total cross section of those reactions
\cite{may96,wil97}.

Based on modern
determinations of the $T$-matrix of $\eta N$ scattering
\cite{bat95,gre97}, theoretical calculations of the $\eta$-nucleus
scattering length $a_{\eta A}$ were made \cite{rak96}, with
the conclusion that $\eta$-nuclei ${}_\eta A$ must exist for all $A \ge 3$.

It should be kept in mind that the purely experimental evidence from
the reactions with $\eta$ in the final state do not determine the {\em
sign} of $a_{\eta A}$ \cite{may96} and thus they cannot unambiguously
prove that $\eta$-nuclei really exist as a bound rather than virtual
state.  Therefore, a crucial experiment would be to observe the
bound $\eta$'s. That is the aim of the present work.
Specifically, we performed a search for $\eta$-nuclei in the reaction
\begin{equation}
    \gamma + {}^{12}\mbox{C} \to N + {}_\eta(A-1)
          \to N + \pi^+ n + (A-2),
\end{equation}
in which were detected decay products of the $\eta$-nuclei, viz.\ the
correlated pions and nucleons emitted in opposite directions
transversely to the beam (Fig.~1). The underlying idea \cite{sok91} is
that such $\pi N$ pairs cannot be produced in quasi-free
photoproduction at energies as high as $E_\gamma \sim 700$ MeV, whereas
they naturally arise due to $\eta$'s stopped and captured in the nucleus.

\begin{figure}[ht]
\centerline{\epsfxsize=0.4\textwidth \epsfbox[70 745 220 805]{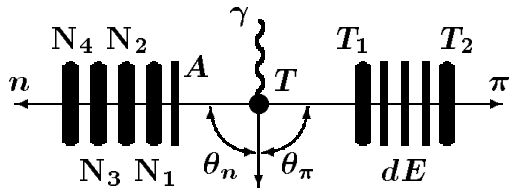}}
\caption{\sl
Layout of the experimental setup. Shown are the time-of-flight
pion and neutron spectrometers.}
\end{figure}

The process of the $\eta$-nucleus formation in the reaction (1)
followed by its decay is shown schematically in Fig.~2a. There, both
the first stage of the reaction, i.e.\  production of $\eta$ by a
photon, and the second stage, i.e.\ annihilation of $\eta$ and creation
of a pion, proceeds through single-nucleon interactions
(either with a proton or a neutron in the nucleus),
mediated by the $S_{11}(1535)$
nucleon resonance.  Formation of the bound state of $\eta$ and the
nucleus becomes possible when the momentum of the produced $\eta$ is
small --- typically less than 150 MeV/c (see Fig.~3).  The
kinematics suggests photon energies $E_\gamma = 650{-}850$ MeV as the most
suitable for creating the $\eta$-nuclei.  Due to the Fermi motion, $\pi N$
pairs from $\eta$-nucleus decays have the characteristic opening angle
$\langle\theta_{\pi N}\rangle = 180^\circ$ with the width of $\simeq
25^\circ$. The kinetic energies are $\langle E_\pi\rangle \simeq 300$
MeV and $\langle E_n\rangle \simeq 100$ MeV.  In the case when the
momentum (or energy) of the produced $\eta$ is high, the attraction
between $\eta$ and the nucleus is not essential, and the $\eta$-meson
propagates freely (up to an absorption, see Fig.~2b).  In this case, the
final $\pi N$ pairs also carry a high momentum and their kinematical
characteristics are different from those of
pairs produced through the stage of the $\eta$-nucleus formation.

\begin{figure}[ht]
\centerline{\epsfxsize=0.7\textwidth \epsfbox[30 700 340 780]{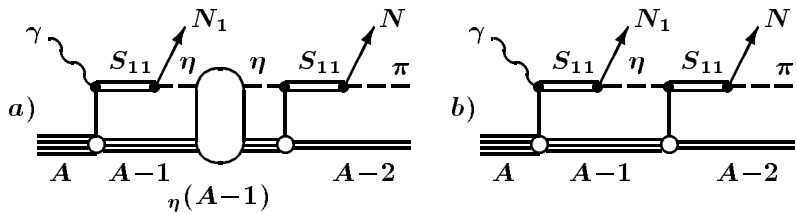}}
\caption{\sl
(a) Mechanism of formation and decay of an $\eta$-nucleus.
(b) Background production and decay of $\eta$'s in the nucleus.}
\end{figure}

A systematic way to describe both the resonance (Fig.~2a) and
background (Fig.~2b) processes consists in using the Green function
$G({\bf r}_1,{\bf r}_2,E)$ which gives an amplitude of $\eta$ having the
energy $E$ to propagate between the creation and annihilation points in
the nuclear mean-field described by an optical energy-dependent
potential $U(r,E)$. In the vicinity of a bound level of a (complex)
energy $E_0$, the Green function has a pole $\sim 1/(E-E_0)$, and this
pole corresponds to the mechanism shown in Fig.~2a.  The
background process (Fig.~2b) corresponds to a non-pole part of
$G$. A convenient measure of the relative role of the background and
resonance processes is given by the spectral function
  $S(E) = \int \!\!\!\int \rho(r_1)
        \,\rho(r_2) \, |G({\bf r}_1,{\bf r}_2,E)|^2
        \,d{\bf r}_1 \, d{\bf r}_2$
(cf.\  \cite{mor85}),
which characterizes a nuclear dependence of pion production through
the two-step transition $\gamma \to \eta \to \pi$ in the nucleus.
$S(E)$ depends on the binding potential $U$ and it is proportional to
the number of $\eta N$ collisions which $\eta$ experiences
when passes through the nucleus of the density $\rho(r)$
between the creation and annihilation points.
The attractive potential $U$ makes the produced $\eta$ of a
near-resonance energy $E$ to pass several times through the nucleus
before it decays or escapes,
thus resulting in an enhanced number of collisions and in a resonance
increasing the production rate of the correlated $\pi N$ pairs.

\begin{figure}[ht]
\centerline{\epsfxsize=\textwidth \epsfbox[50 50 540 245]{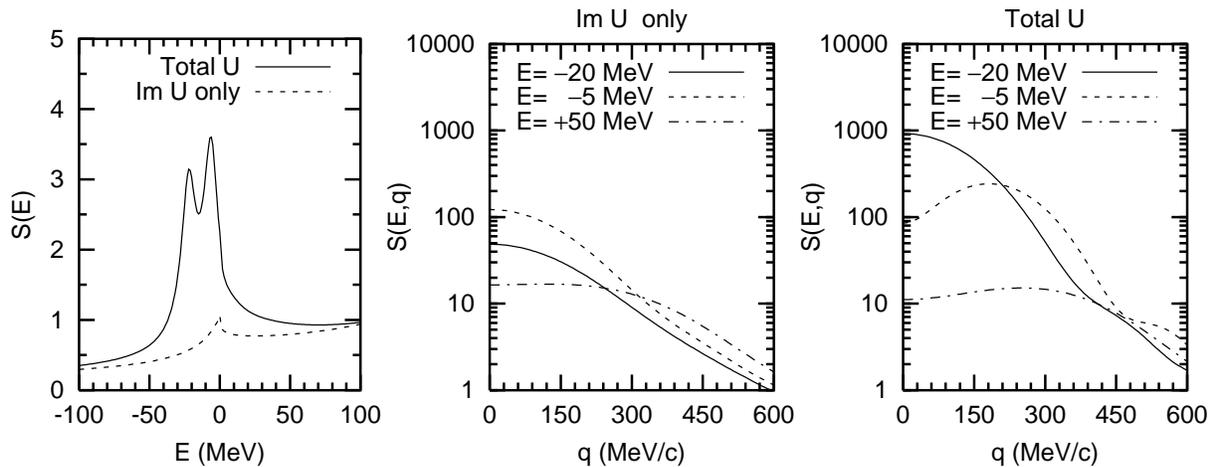}}
\caption{\sl
Spectral functions $S(E)$ and $S(E,q)$ (in arbitrary units)
calculated for a rectangular-well optical potential
simulating the nucleus $^{12}$C.
For a comparison, the results obtained by dropping out the attractive
(i.e.\ real) part of the $\eta A$ potential $U$ are also shown.}
\end{figure}

A comparative role of the resonance and background contributions is
illustrated in Fig.~3 \cite{lvo98}, in which the spectral function
$S(E)$ is shown for the case of a rectangular-well optical potential $U$
simulating the $^{12}$C nuclear density which is proportional to
the elementary $\eta N$-scattering amplitude
\cite{gre97}. The $\eta A$ attraction results in a prominent enhancement in
the number of collisions when $\eta$ has a negative energy between
0 and $-30$ MeV.
The related spectral function $S(E,q)$
is given by the Fourier components of the inner part
of the Green function
(viz.\ a part having an overlap with nucleons in the nucleus), and it
describes a nuclear dependence of the energy-momentum
distribution $\partial^2 N / \partial E \partial{\bf q}$ of
the produced $\pi N$ pairs over their total energy and momentum
(which are $E + m_\eta + m_N$ and $q$, respectively,
up to the Fermi smearing).
As seen in Fig.~3, the $\eta$-nucleus attraction results
in a strong enhancement in the momentum density
at the resonance energies $E$ and low $q$.
This theoretical finding supports the starting point of the further
analysis that the correlated $\pi N$ pairs predominantly appear
from decays of bound $\eta$'s.

The experimental setup (Fig.~1) consisted of two scintillation
time-of-flight spectrometers having the apparatus time-resolution
of $\Delta\tau=50$ ps.  Both spectrometers were
positioned around a 4 cm carbon target $T$ at either $\theta=50^\circ$
or $90^\circ$ with respect to the photon beam (on its opposite sides),
each covering a solid angle of $\sim 0.06$ sr.
An anticoincidence-counter ($A$)
of charged particles located in front of the neutron detectors had the
rejection efficiency of 90\%.

\begin{figure}[ht]
\centerline{\epsfxsize=1.1\textwidth \epsfbox[23 450 470 790]{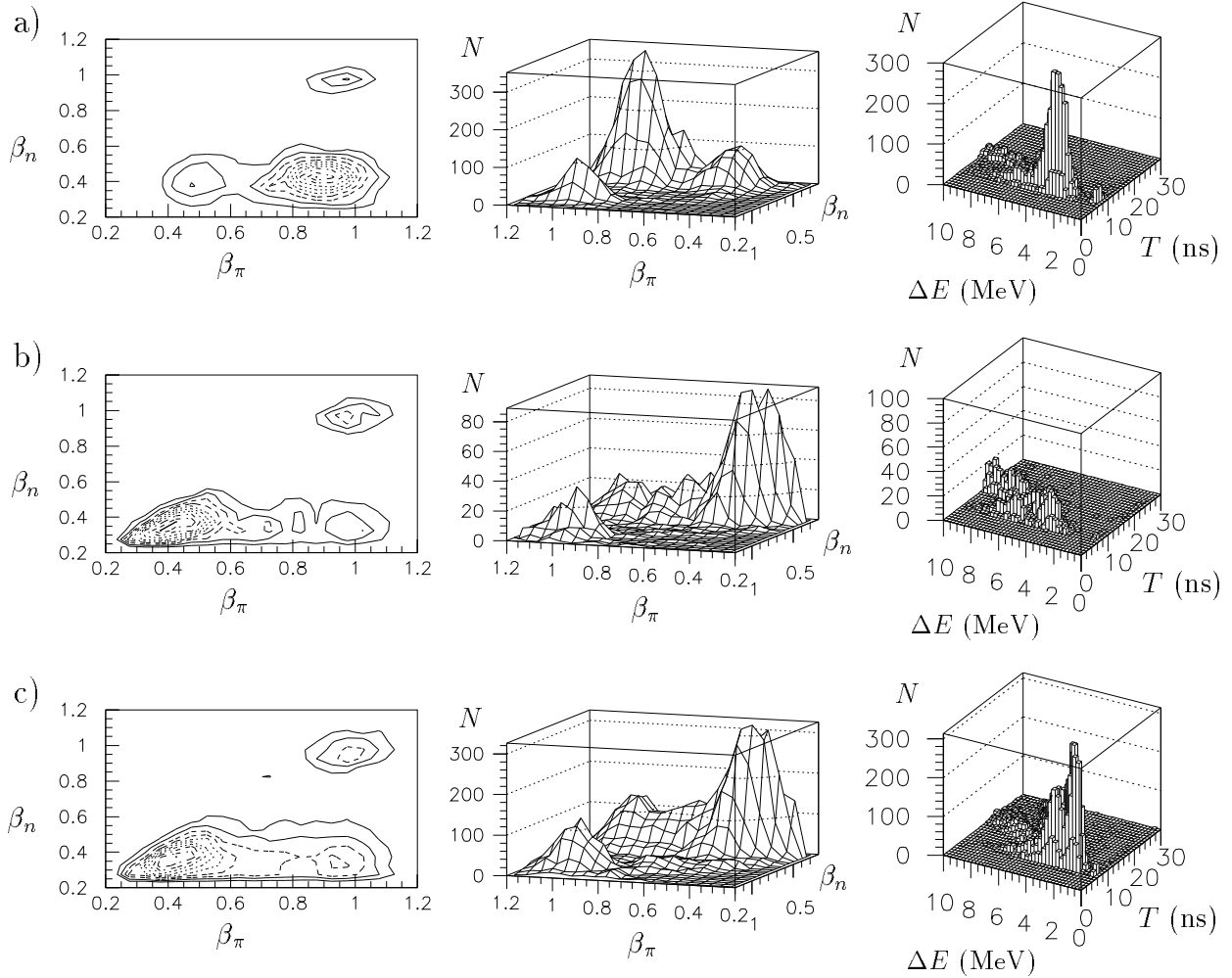}}
\caption{\sl
Left and central panels:  Distributions (the number of events
$N$) over the pion and neutron velocities for (a) the ``calibration",
(b) ``background", and (c) ``effect$+$background" runs.  Right panels:
Distributions over the time-of-flight $T$ and the energy losses $\Delta
E$ in the pion spectrometer for the same runs; only events with a slow
particle in the neutron spectrometer were selected for these plots.}
\end{figure}

Three runs were made in the present experiment with different
positions $\theta=\theta_n=\theta_\pi$ of the spectrometers:
(a) a ``calibration", (b) ``background", and
(c) ``effect $+$ background" runs.  In the
``calibration" run (a), both spectrometers were placed at
$\theta=50^\circ$ with respect to the photon beam, and the end-point
energy of the bremsstrahlung spectrum was $E_{\gamma \rm max}=650$ MeV.
In this run, mainly $\pi^+n$ pairs from quasi-free production of pions
on the carbon, $\gamma + {}^{12}\mbox{C} \to \pi^+ +  n + X$, were
detected.  In the ''background" run (b), the spectrometers were moved
to $\theta=90^\circ$ with respect to the photon beam, i.e.\ to the
position suitable for measuring the effect. The end-point
energy was still $E_{\gamma \rm max}=650$ MeV, i.e.\  well below the
$\eta$ photoproduction threshold off free nucleons which is 707 MeV.
In the "effect$+$background" run (c), keeping the angle
$\theta=90^\circ$, the beam energy was set above the threshold:
$E_{\gamma \rm max}=850$ MeV.  The observed two-dimensional velocity
spectra of the detected pairs are shown in Fig.~4 for all three runs.

According to the velocities of detected particles in the pion and
neutron spectrometer, all events in each run were divided into
three groups:  fast-fast (FF), fast-slow (FS), and slow-slow (SS).  The
FF events with the extreme velocities close to the speed of the light
correspond to a background (mainly $e^+e^-$ pairs produced by $\pi^0$
from double-pion production).  The FS events mostly correspond to $\pi
N$ pairs.  In the ``calibration" run ($\theta=50^\circ$, $E_{\gamma \rm
max}=650$ MeV), the quasi-free production of the $\pi^+n$ pairs is seen
as a prominent peak in the two-dimensional distribution (Fig.~4a).  In
the ``background" run ($\theta=90^\circ$, $E_{\gamma \rm max}=650$
MeV), the largest peak (SS events in Fig.~4b) is caused by $\pi\pi$
pairs from double-pion photoproduction off the nucleus.  In the
``effect$+$background" run ($\theta=90^\circ$ and $E_{\gamma \rm
max}=850$ MeV) (Fig.~4c), apart from the SS events, a clear excess of
the FS events, as compared with the ``background" run, is seen.  This
FS signal is interpreted as a result of production and annihilation of
slow $\eta$'s in the nucleus giving the $\pi^+n$ pairs.

A further analysis of the events was done by using an information from
three scintillation detectors which were positioned between the
starting and finishing layers, $T_1$ and $T_2$, of the time-of-flight
pion spectrometer (Fig.~1); they measured the energy losses $\Delta E$
of particles that passed through them.
A selection of events with a minimal
$\Delta E$ in the two-dimensional distributions over the time of flight $T$
and the energy losses $\Delta E$ (Fig.~4) allows to discriminate events
with a single pion from those with the $e^+e^-$ pairs.

The counting rate of the $\pi^+n$ events was evaluated as
\begin{equation}
  N(\pi^+n; 850) =   N({\rm FS_{min}}; 850) - N({\rm FS_{min}}; 650)
    \times K(850/650),
\end{equation}
where $N({\rm FS_{min}};E_{\gamma \rm max})$ is the number of the
observed FS events with the minimal $\Delta E$ and with the specific
photon energy $E_{\gamma \rm max}$, and where the coefficient K(850/650)
gives an increase of the FS-background due to double-pion
photoproduction when $E_{\gamma \rm max}$ is increased
from 650 MeV up to 850
MeV.  Assuming that the same coefficient describes also an increase of the
SS count rate, it was found from the SS events at 650 and 850
MeV that $K=2.15$.  Such a procedure gives $N(\pi^+n; 850) = (61 \pm
7)$ events/hour.  Assuming an isotropic distribution of the $\pi^+n$
pairs, taking into account efficiencies of the pion and neutron
spectrometers (80\% and 30\%, respectively) and evaluating a
geometrical fraction $f$ of the correlated $\pi^+n$ pairs
simultaneously detected by the pion and neutron detectors of a finite
size ($f=0.18$ was determined by a Monte Carlo
simulation of the width of the angular correlation between $\pi^+$ and
$n$ caused by the Fermi motion of nucleons and $\eta$ in the nucleus), we
obtain the following estimate of the total photoproduction cross
section of the correlated pairs from carbon averaged over the
energy interval of 650--850 MeV:~
$  \langle \sigma(\pi^+n) \rangle =  (12.2 \pm 1.3) ~\mu\mbox{b} $
(a statistical error only).

Summarizing, we have observed a clear excess of the correlated $\pi^+n$
pairs with the opening angle close to $180^\circ$ arising when the
photon beam energy becomes higher than the $\eta$-production threshold.
That is, we have observed production and decay of slow $\eta$'s inside
the nucleus.  As was discussed above when describing Fig.~3,
these pairs are expected to be mostly related with a formation
and decay of $\eta$-nuclei in the intermediate state.
The obtained total cross section of pair production
is close to the theoretical predictions \cite{try95} for the total
cross section of $\eta$-nuclei formation in photo-reactions,
what provides a further support for that expectation.
For more direct arguments in favor of the observation of $\eta$-nuclei,
angular and energy distributions of the components of the pairs
have to be analyzed and a better statistics has to be achieved.
This work is in progress now.

\bigskip
\centerline{\it Acknowledgements}
\bigskip

The present research was supported by the Russian Foundation for
Basic Research, grant 96-02-17103.

\end{document}